\documentclass[conference]{IEEEtran}
\IEEEoverridecommandlockouts
\usepackage[pdftex]{graphicx}
\usepackage{epstopdf}
\usepackage{psfrag,graphicx}
\usepackage{verbatim}
\usepackage{amsfonts}
\usepackage{caption}
\usepackage{amssymb}
\usepackage{amsmath}
\usepackage{amsthm}
\usepackage{subfigure}
\usepackage{subfig}
\usepackage{graphicx} 
\usepackage{color}
\usepackage[ruled,vlined]{algorithm2e}

\begin{document}

\title{Blind Null-space Tracking for MIMO Underlay Cognitive Radio Networks}

\author{\authorblockN{Alexandros Manolakos, $  $    Yair Noam $  $ and $  $   Andrea J. Goldsmith} \authorblockA{Department of Electrical  Engineering\\ Stanford University, Stanford CA 94305\\ Email: \{amanolak,noamyair,andreag\}@stanford.edu}  }

\maketitle 

\begin{abstract}
Blind Null Space Learning (BNSL) \cite{Noam}  has recently been proposed for fast and accurate learning of the null-space associated with the channel matrix   between a secondary transmitter and a primary receiver. In this paper we propose a channel tracking enhancement  of the algorithm, namely the Blind Null Space Tracking (BNST) algorithm that allows  transmission of information to the Secondary Receiver (SR) while simultaneously learning the null-space of the time-varying target channel. Specifically, the enhanced algorithm initially performs a BNSL sweep  in order to acquire the null space. Then, it performs modified Jacobi rotations such that the induced interference to the primary receiver is kept lower than a given threshold $P_{Th}$ with probability $p$ while information is transmitted to the SR simultaneously. We present  simulation results indicating that the proposed approach has strictly better performance over the BNSL algorithm for channels with independent Rayleigh fading with a small Doppler frequency. 
\end{abstract}

\section{Introduction}

MIMO communication has reformed the way we think about wireless communications. in particular, MIMO can be  revolutionary in the paradigm of the underlay cognitive radios. The motivation of the underlay cognitive radios is to allow cognitive users to operate simultaneously in the same frequency band as primary users while causing them minimal interference (\cite{goldsmith2}, \cite{Haykin}).

The starting point of this work is the recent publications from Noam and Goldsmith (\cite{Noam}, \cite{Noam2}) where the authors provide a null space learning algorithm with which a MIMO secondary transmitter (SU-Tx) can learn the null space of the interference channel in order not to interfere with the primary receiver (PU-Rx). The algorithm has  nice convergence properties, it can work  with only energy measurements, and it is independent of the modulation and other transmission parameters of the primary system.

Typical channel estimation problems are divided into two categories: channel acquisition and channel tracking (\cite{goldsmith}).  Channel acquisition refers to the problem of learning the channel without any prior knowledge. Channel tracking refers to the problem of  continually updating the channel estimate, using a previous estimate.   {  A solution to the channel acquisition problem  is provided in [3] under the assumptions of reciprocity of the channels, TDD transmission to both the primary and secondary systems, and a  transmission frame structure of the secondary system that is synchronized with that of the primary system.}

In the case of time-varying channels, obtaining the null-space only once (i.e. null space acquisition), as it is performed in $\cite{Noam}$, is only the first step; a constant tracking algorithm that updates the null space continuously is needed. Unlike channel acquisition,  channel tracking needs to be combined with simultaneous transmission of information since it is not a one time event, otherwise there will be a non-negligible reduction in data rate. Therefore, the channel tracking algorithm must ensure that the interference induced to the primary network remains low with high probability at all times.  We introduce such a tracking algorithm  to learn the channel and send information at the same time to the SR. 

The remainder of this paper is organized as follows: In Subsection \ref{model2} we present the channel model, in  \ref{BNSL} we describe briefly the BNSL algorithm and in \ref{Metrics} we indicate the performance metric that is used to test the tracking algorithm. Then, Section \ref{variations} presents simulation results that indicate that the null-space of a MIMO Rayleigh fading channel varies faster than the channel coherence time indicates. Section \ref{tracking} presents the Blind Null Space Tracking algorithm (BNST) and compares its performance with the BNSL algorithm. Moreover, in Section \ref{dream} we describe and analyze a general transmission scheme of information that can be used along with the BNST algorithm. Simulation results of the BER performance of this scheme are presented in Section \ref{Simulation}. Finally, Section \ref{Conclusions} concludes the paper and discusses future research directions.
 
 \section{System Model}

 \subsection{Channel Model}
 \label{model2}
\begin{figure}[h!]
\begin{center}
\includegraphics[width=0.6\columnwidth]{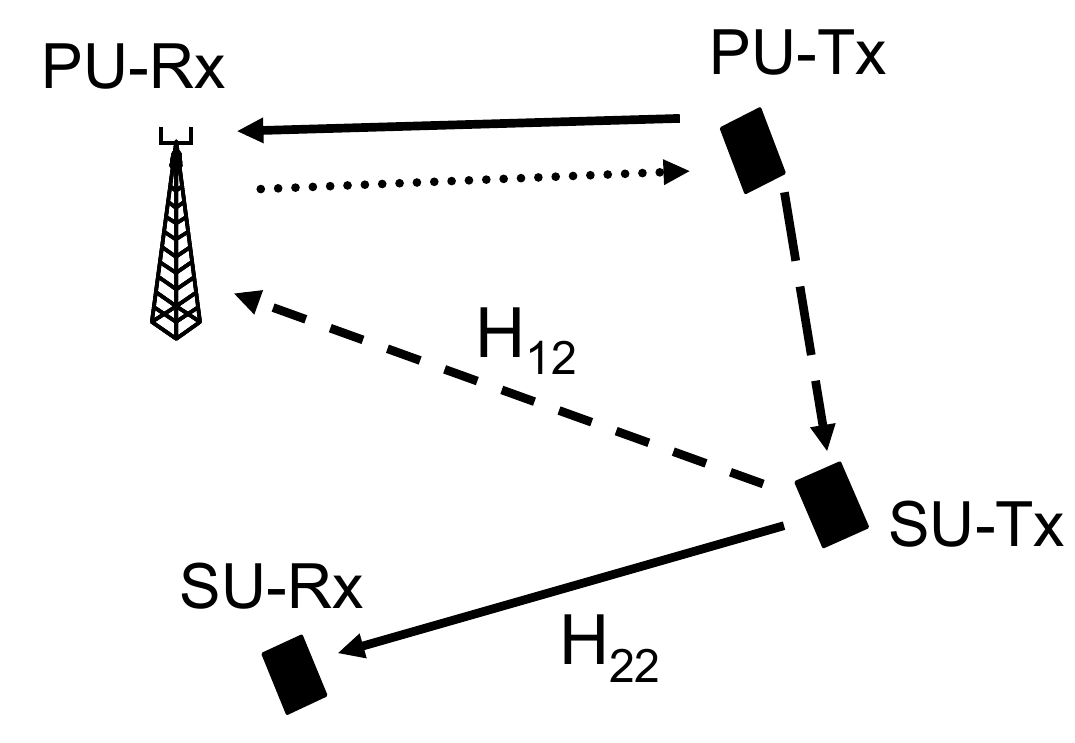}
 \caption{{  The addressed cognitive radio scheme. \label{figure1}}}
\end{center}
\end{figure}

Consider the cognitive radio scenario depicted in Figure \ref{figure1} in which a Secondary User transmitter (SU-Tx)  is allowed to coexist with a Primary User receiver (PU-Rx) as long as the interference it inflicts to him is below some threshold. Denote as $(N_t,N_r)$-system a system with $N_t$ antennas in the SU-Tx and $N_r$ antennas in the PU-Rx.  {  To mitigate the interference to the PU-Rx, the SU-Tx    would like to learn the interference channel matrix  between the SU-Tx and the PU-Rx denoted as $\mathbf{H}_{12}(t) =[h_{ij}(t)] \in \mathbb{C}^{N_r \times N_t}$, where $h_{ij}(t)$ represents  the channel between the  SU-Tx's $j\mbox{-th}$ antenna and the  PU-Rx's  $i\mbox{-th}$ antenna. }

Denote as $T_s$ the symbol duration, as $F_c$ the carrier frequency, as $\mathbf{T}(t) \in \mathbb{C}^{N_t \times (N_t-N_r)}$ the precoding matrix that the SU-Tx uses and as $\mathbf{H}_{22}(t)$ the channel matrix between SU-Tx and SU-Rx.  We make the assumption of Wide Sense Stationary Uncorrelated Scattering (WSSUS) and that the channels $h_{ij}(t)$ have independent Rayleigh fading with maximum doppler frequency $F_d$.  This channel represents the most difficult channel to track since there is no correlation among the $h_{ij}(t)$ and no line-of-sight (LOS) component. We also assume that  between one cycle and another, the variation that the SU-Tx is making to its signal is the dominant factor that varies the PU-Rx's SINR. Lastly, throughout this paper we consider only flat fading channels.
\subsection{Blind Null Space Learing Algorithm}
\label{BNSL}

We now present the BNSL algorithm, for null space acquisition, from which we derive the null space tracking algorithm (BNST) that is presented in Section \ref{tracking}. 

Assume $\mathbf{H}_{12}$ is constant and denote $\mathbf{G} = \mathbf{H}_{12}^*\mathbf{H}_{12}$, whose null-space is the same as $\mathbf{H}_{12}$. The leaning process is carried out in cycles during which the SU-Tx transmits a learning signal that affects the PU-Rx interference level. We assume that the primary system is using a transmission scheme which adapts the  transmitted signal according to the SINR at the PU-Rx. We further  assume that the SU-Tx can sense whether the interference it inflicts to the PU-Rx increases or decreases from listening the PU-Tx transmitted signal or control channel. Denote as $T_{FB}$ the amount of time needed from the time that the SU-Tx starts transmitting a learning message until it  measures the  effect of its learning signal on the PU-Tx's transmit power. Using this knowledge, the SU-Tx tries to ``stir'' the beam forming pattern of its transmissions such that it causes minimal interference to the PU-Rx, i.e. it transmits in the null space of $\mathbf{H}_{12}$.

The learning of the null-space is carried out in sweeps. Each sweep consists of $N_t(N_t-1)/2$  learning stages where each one consists of $K$ transmission cycles of length of $T_{FB}$ msec. The SU-Tx sends constant learning signals  inside each transmission cycle, denoted as $\mathbf{\tilde{x}}(n)$ where $n$ denotes the index of the current transmission cycle. The heart of the ``rotation'' mechanism is the Jacobi Eigenvalue Decomposition. Specifically, the Jacobi technique obtains the eigenvalue decomposition of the Hermitian matrix $\mathbf{G}$ via a series of 2-dimensional rotations that eliminate two off-diagonal elements of $\mathbf{G}$  in each learning stage. The algorithm begins by setting $\mathbf{A}_0 = \mathbf{G}$ and then performs the following rotation operations $\mathbf{A}_{k+1} = \mathbf{R}_{l_k,m_k}(\theta,\phi) \mathbf{A}_k \mathbf{R}_{l_k,m_k}^*(\theta,\phi)$ where $\mathbf{R}_{l_k,m_k}^*(\theta,\phi)$ is an $N_t\times N_t$ rotation matrix who is equal to the identity matrix except for its $(l_k,l_k)$ and $(m_k,m_k)$ entries, which are equal to $cos(\theta)$ and its $(l_k,m_k)$ and $(m_k,l_k)$ entries, which are equal to $e^{-i\phi}\sin(\theta)$ and  $-e^{-i\phi}\sin(\theta)$  respectively. The rotation angles $\theta, \phi$ are chosen to zero the $(l_k,m_k)$ entry of $\mathbf{A}_k$, i.e., $[\mathbf{A}_k]_{l_k,m_k}$.  The integers $l_k,m_k$ are the indices of the off-diagonal entries that are eliminated in the $k$-th step and are chosen such that the off-diagonal elements are eliminated periodically, where each  entry is eliminated once in each learning stage [1]. Noam and Goldsmith proposed a blind technique (without observing the matrices $A_k, k=0,1,\dots$) in which   the SU-Tx obtains       $\theta$ and $\phi$ with only two binary line searches. In other words, the SU-Tx in each learning stage $k$ calculates a pre-coding matrix $\mathbf{W}_k$ such that $\mathbf{W}_k^{*}\mathbf{G}\mathbf{W}_k=\mathbf{A}_k$. To do so, it  samples the $\theta-\phi$ plane and gets feedback from sensing the transmit power of the PU-Tx. After $N_t(N_t-1)/2$ learning stages , which constitute one Jacobi sweep, the SU-Tx has the first estimate of the eigenspace.

\subsection{Performance Metric}
\label{Metrics}
In order to compare the different null space learning methods we need to introduce a natural performance metric. The metric that the primary system is concerned about is the maximum interference caused by the SU-Tx.  Denote as $\mathbf{x}(t)$ the message that the latter wants to send. Then, the normalized  interference is upper bounded by  the spectral norm of the matrix $\mathbf{H}_{12}(t) \mathbf{T}(t)$ as:
\begin{align}
||\mathbf{H}_{12}(t) \mathbf{T}(t) || =\max\limits_{||\mathbf{x}(t)||_2 \neq 0} \frac{||\mathbf{H}_{12}(t)  \mathbf{T}(t) \mathbf{x}(t) ||_2}{|| \mathbf{x}(t)||_2}
\label{norm}
\end{align}
\noindent We  say that the null-space tracking is successful  if
\[
\mbox{Pr} \left \{ 10 \log_{10} \frac{ ||\mathbf{H}_{12}(t) \mathbf{T}(t) ||^2}{||\mathbf{H}_{12}(t) ||^2}\leq - P_{Th} \right \}  = p,
\]
\noindent that is, if $p$ percent of the time  the interference achieved with the use of the precoding matrix $\mathbf{T}(t)$ is at least $P_{Th}$ dB less than if we use as $\mathbf{T}(t) = \mathbf{I}_{N_t \times N_t}$. 

\section{Null space coherence time}
\label{variations}

In this section we motivate the need for using the BNST algorithm based on the observation that even in slowly Rayleigh fading MIMO channels, the null space changes much faster than the channel coherence time associated with each of the independent channels $h_{ij}(t)$ indicates.

The coherence time $T_c$ is a widely accepted  characteristic of a channel that is used to quantify how fast it changes. We assume that the correlations between channel coefficients at different times depends only on the time difference $\Delta$t. We define the  normalized autocorrelation of a SISO channel $h(t)$ as (chapter 3.3 \cite{goldsmith})
\begin{align}
\rho_{\Delta t} = \frac{E\{h(t)h(t+\Delta t)^*\}}{E\{|h(t)|^2\}}
\end{align}

\noindent Then the $X\%$ coherence time, denoted as $T_c^{[X]}$, is defined as the value $\Delta t$ such that $\rho_{\Delta t} = \frac{X}{100}$, i.e. $\rho_{T_c^{[X]}} = \frac{X}{100}$.  For example, in Clarke's model \cite{goldsmith}, $T_c^{[0.5]} = \frac{9}{16\pi F_d}$.

\begin{figure}[h!]
\begin{center}
\includegraphics[width=0.8\columnwidth]{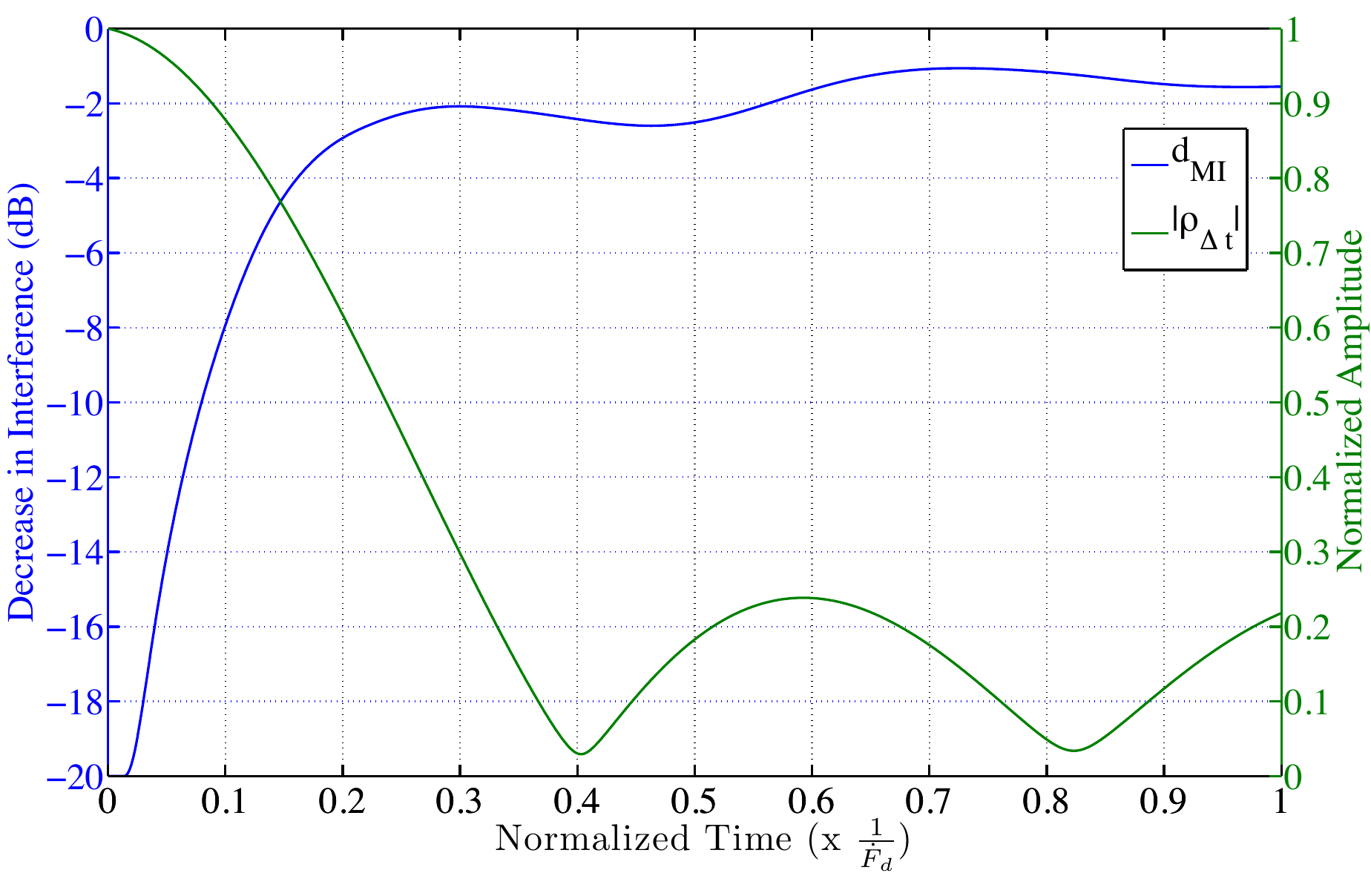}
 \caption{ $(3,1)$-system. The left vertical axes represents $d_{MI}(\Delta t)$ and the right vertical axes $\rho_{\Delta t}$. Horizontal axes represents the normalized time as a a fraction of $\frac{1}{F_d}$\label{fig4}.}
\end{center}
\end{figure}

To study the relation between the $\mathbf{H}_{12}(t)$'s null space time variations and the coherence time of its entries, we turn to simulations. Consider a $(3,1)$-system  where  each $h_{ij}(t)$ is an  independent Rayleigh fading channels (i.e. the entries of $\mathbf{H}_{12}(t)$ are independent of each other) with Doppler Frequency  of $F_d=6.48~Hz$ (e.g. relative speed of 10 Km/h and carrier frequency of $F_c = 700~Mhz$)  using the Clarke's model with 40 multipath components. Figure \ref{fig4} depicts the amplitude of the normalized autocorrelation $|\rho_{\Delta t}|$ and the relative interference reduction  $d_{MI}(\Delta t)$ i.e.

\begin{align}
d_{MI}(\Delta t) = 10 \log_{10} \left( E \left \{ \frac{ ||\mathbf{H}_{12}(t+\Delta t) \mathcal{N}(\mathbf{H}_{12}(t)) ||}{||\mathbf{H}_{12}(t+\Delta t) ||}\right \} \right)  
\end{align}

\noindent The latter calculates the average minimum decrease of interference inflicted to the PU-Rx if the current channel is $\mathbf{H}_{12}(t+\Delta t)$ whereas the SU-Tx uses the null space of an outdated channel, i.e. $\mathcal{N}(\mathbf{H}_{12}(t))$. Note that $d_{MI}(\Delta t)$ increases surprisingly fast. For example, $d_{MI} \geq -15$dB for $\rho_{\Delta t} \leq 0.95$! This means that even if the SU-Tx obtains the null space of the channel instantly at time $t$, it will have to update it after $T^{[0.95]}$ seconds, if it needs to cause a minimum of $-15~dB$ interference to the primary system. The intuition behind this observation  is  that $\mathbf{H}_{12}(t)$ is a time-varying matrix whose elements change approximately every $T_c$ seconds. The null space is a quantity that depends on all the elements of the matrix and therefore it is likely to change much faster. This observation was verified using a different Rayleigh fading generator (\cite{Baddour}) and with different numbers of multipath components (from 5 up to 100).

\section{Blind Null Space Tracking}
\label{tracking}

In this section, we summarize the Blind Null Space Tracking (BNST) algorithm and present simulation results which show that in the case of slow-varying MIMO channels it successfully tracks the channel. We postpone the discussion on  how the SU-Tx performs simultaneous transmissions while it tracks the null space in Section \ref{dream}.
 
Denote as $T$ the time needed for a Jacobi sweep to finish. Assume at time $t$ we have an approximation $\mathbf{\tilde{W}}(t)$ of the eigenspace  $\mathbf{W}(t)$ of the channel matrix. 

Consider the Modified Jacobi Sweep that is described in Algorithm \ref{alg1}. Function {\bf NextElement} returns the values $l_k, m_k$ for the $k$-th learning stage. The input is the current estimated eigenspace  $\mathbf{\tilde{W}}(t)$ along with the parameters $\tilde{\theta},\theta_{max},\eta$ that define the search space of the line searches of the sweep. Specifically,  $\theta_{max}$ denotes that the SU-Tx should perform a line search in the interval $[-\theta_{max},\theta_{max}]$. $\tilde{\theta}$ is the value of $\theta$  used in the {\bf LineSearch} in Line 4 of Algorithm \ref{alg1} and $\eta$ is the step size of the binary search which may vary as a function of $\theta_{max}$.

The channel tracking works as follows. Firstly, the SU-Tx performs one complete sweep of the BNSL algorithm, i.e. it uses $\mathbf{\tilde{W}}(t) = \mathbf{I}_{N_t \times N_t},\tilde{\theta}=\frac{\pi}{3},\theta_{max} =\frac{\pi}{2},\eta=0.1 $. These are the parameters used in \cite{Noam}. This sweep creates the first estimate $\mathbf{\tilde{W}}(t+T)$ of the eigenspace and completes the channel acquisition phase. Then, the SU-Tx transmits information using as the precoding matrix $\mathbf{T} = [\mathbf{\tilde{w}}_{1},\mathbf{\tilde{w}}_{1},\dots,\mathbf{\tilde{w}}_{N_t-N_r}$], where $\mathbf{\tilde{w}}_i$ are the columns of the matrix $\mathbf{\tilde{W}}(t+T)$, i.e. the SU pre-multiplies its transmitted signal $\mathbf{t+T}$ by the matrix $\mathbf{T}$.  The modified BNSL sweep always returns the $\mathbf{\tilde{W}}(t+T)$ such that the singular values that correspond to  its column are in a decreasing order, i.e. let $\sigma_i$ be the singular value that corresponds to $\mathbf{\tilde{w}}_i$, then $\sigma_i\leq \sigma _{i+1},\; 1\leq i<N_t$. 

The SU-Tx continuously senses the transmit power of the PU-Rx. When it senses that the transmit power has increased more than a predefined threshold, then it updates the null space's  estimate. In our current simulations we make the assumption that the SU-Tx knows \footnote{In \cite{Noam}, it is shown how can the  SU  autonomously estimate the interference reduction without cooperation from the primary user } the  quantity $||\mathbf{H}_{12}(t) \mathbf{T} ||_{dB}$ and it performs an adaptation whenever $||\mathbf{H}_{12}(t) \mathbf{T} ||_{dB} < P_{Tr}$ where $P_{Tr}=-20$. The intuition behind this rule is that the SU-Tx should not  wait until the current eigenspace has changed significantly since in that case the matrix $\mathbf{\tilde{W}}(t+T)$ cannot be of much help in the subsequent sweep.

In order to update fast, accurately and with minimal interference caused to the primary system, the algorithm performs  a local search over the space of the $\theta$ angles. This local search should be performed for the following two reasons.  Firstly, each line search can be much faster, assuming that the SU-Tx keeps the same parameter $\eta$. Secondly, given that the SU-Tx has a good estimate of the null space, there exists an explicit trade-off between how large  $\theta_{max}$ should be and the inflicted interference to the primary system . To be precise, it is known that if the estimate is very close to the real null space then $\hat{\theta} \approx 0$ (\cite{Noam}), i.e. there is no need of a ``large rotation''. Therefore, the SU-Tx should not search for the optimal $\theta$  over a large interval around $0$ since it knows that it has a good approximation already. A similar reasoning explains why $\tilde{\theta}$ is an important parameter of the BNST; the SU-Tx should perform the first line search (Line 4 in Algorithm \ref{alg1}) while it is transmitting close to the approximated null space.  If the SU-Tx has a good estimate of the eigenspace, a small $\tilde{\theta}$ means that the line search in line 4 of the modified rotation algorithm is performed around the null space of the channel. Then, the SU-Tx manages to adapt and still keep the inflicted interference in low levels. 

Note that if the estimate is severely outdated, either because the SU-Tx waited too long to perform another sweep, or because the doppler frequency is large, then the SU-Tx should search for the best $\theta$ over the whole interval $[-\pi/2, \pi/2]$. In our simulations  $\theta_{max}$ was  $\frac{\pi}{10}$ and  $\frac{\pi}{5}$ for small ([$1-2~Hz$]) and large ($>2~Hz$) doppler frequencies respectively. A complete sweep uses  $\theta_{max} = \frac{\pi}{2}$. The parameters $\tilde{\theta},\theta_{max}$ should be chosen as a function of the doppler frequency $F_d$. These can be predefined quantities that the SU-Tx chooses after it has estimated the doppler frequency of the channel.


\begin{algorithm}[!t]
\KwIn{ $\tilde{W}(t),\tilde{\theta},\theta_{max},\eta$ } 
\KwOut{ $\mathbf{\tilde{W}}(t+T)$ }
{1. $k=1$\;}
{2. \textbf{while} $k \leq\frac{N_t(N_t-1)}{2}$}\\
{3.~~~~~($l_k,m_k$) = \textbf{NextElement}(k)}\\
{4.~~~~~\mbox{Define} $w_1(\phi) = h(\mathbf{\tilde{W}}(t) \cdot \mathbf{r_{l_k,m_k}(\tilde{\theta},\phi)})$}\\
{5.~~~~~\mbox{Perform} $\hat{\phi} = \mbox{\textbf{LineSearch}}(w_1(\phi),\pi)$}\\ 
{6.~~~~~\mbox{Define} $w_2(\theta) = h(\mathbf{\tilde{W}}(t) \cdot \mathbf{r_{l_k,m_k}(\theta,\hat{\phi})})$}\\
{ 7.~~~~~\mbox{Perform} $\hat{\theta} = \mbox{\textbf{LineSearch}}(w_2(\theta),\theta_{max})$}\\
{ 8.~~~~~\mbox{Set} $\mathbf{\tilde{W}}(t)=\mathbf{\tilde{W}}(t) \cdot \mathbf{R}_{l_k,m_k}(\hat{\theta},\hat{\phi} )$} \\
{ 9.~~~~~$k=k+1$} \\
{ 10. $\mathbf{\tilde{W}}(t+T)=\mathbf{\tilde{W}}(t)$}\\
\caption{Modified Jacobi Sweep}
\label{alg1}
\end{algorithm}



\begin{figure}[h!]
\begin{center}
\includegraphics[width=0.5\columnwidth]{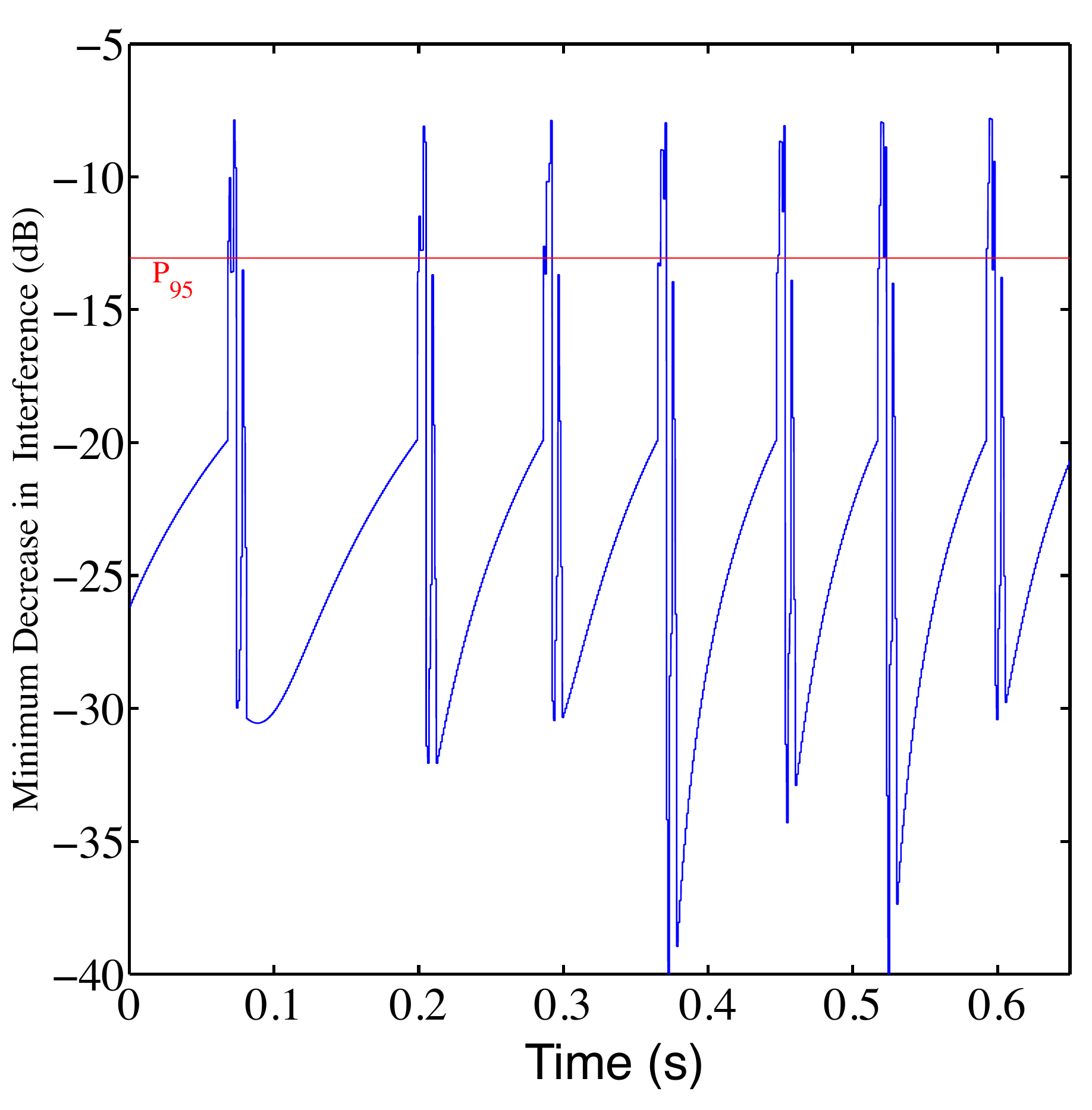}
 \caption{Example for a $(2,1)$-system with $F_d = 1.3$ Hz. \label{fig7}}
\end{center}
\end{figure}

In Figure \ref{fig7}  we show an example ($N_t=2$, $N_r=1$, $T_s = 66.7~\mu$sec, $T_{Fb}=1~$msec \footnote{These parameters are in agreement with the LTE current specifications. Any decrease in $T_{FB}$ will lead to a more accurate and fast tracking.}) of the minimum decrease in interference (in dB) as a function of time when $\mathbf{H}_{12}(t)$ is a  time varying channel with $F_d=1.3$ Hz. 

\begin{figure}[h!]
\begin{center}
\includegraphics[width=0.8\columnwidth]{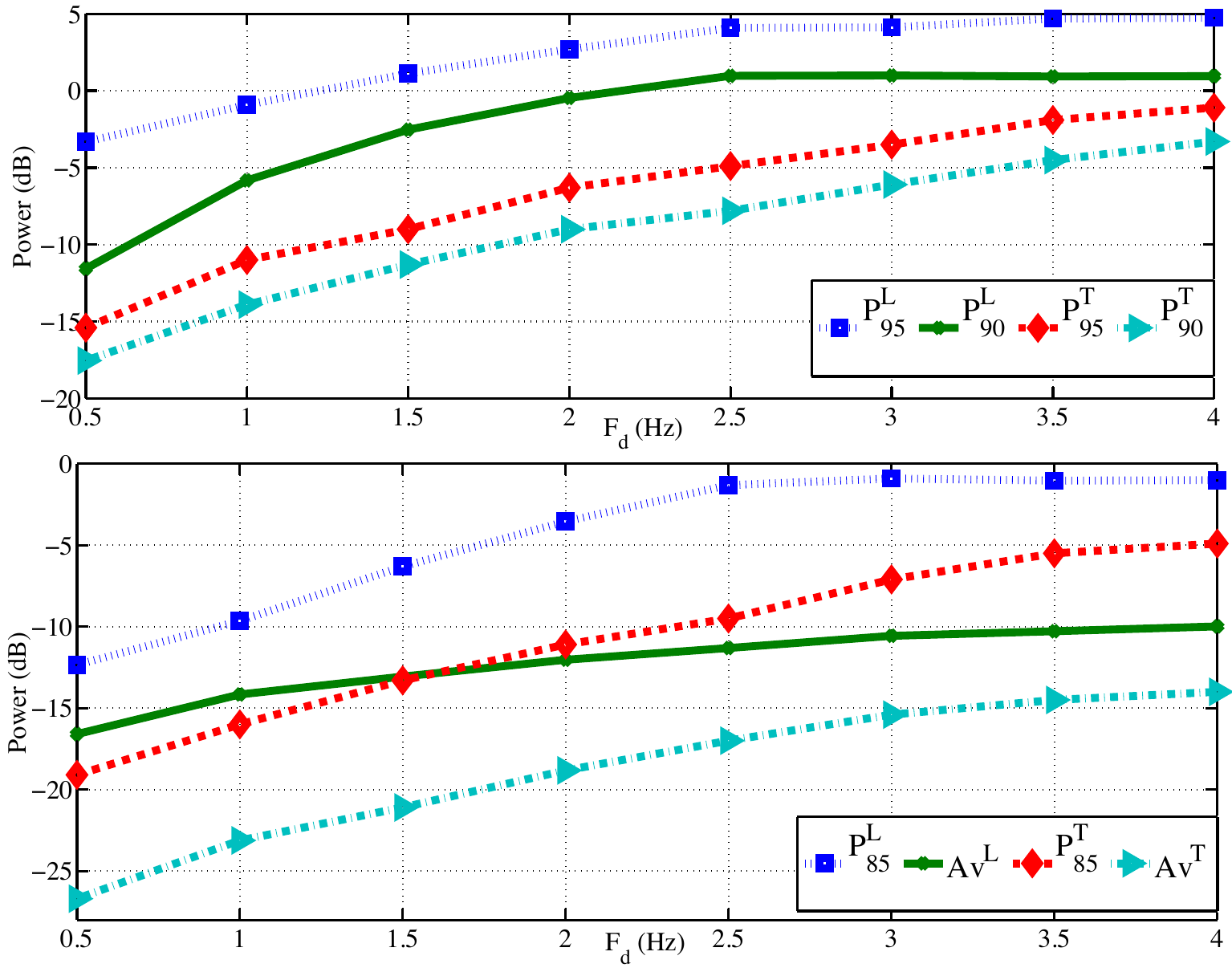}
 \caption{  $P_{95}$, $P_{90}$, $P_{85}$ and Average decrease in interference (Label: Av) for the BNSL and BNST algorithms  for different  $F_d$. Superscript $L$ and $T$ mean BNSL and BNST respectively. \label{finalfig}}
\end{center}
\end{figure}

In Figure \ref{finalfig} we compare the performance of the BNSL algorithm against the performance of the BNST algorithm over a $(2,1)$-system. We average over $100$ different channels and $10^4$ slots for each channel for different doppler frequencies $F_d$. The performance metrics that we use are $P_{95}$, $P_{90}$, $P_{85}$ and Average Decrease of Interference, where $P_{X}$ represents the   decrease of interference (in dB) in $X\%$ percentage of the slots compared to the system that does not use any precoding matrix. For instance, if $P_{90}=-10$ dB it means that in $90\%$ of the slots the SU-Tx caused at least $-10$ dB less interference by using the proposed algorithm.  Note that whenever the SU-Tx adapts we calculate the actual interference caused to the PU-Rx since the SU-Tx sends known signals defined from the algorithms. On the other hand, when the SU-Tx is not adapting we use the maximum interference as defined in (\ref{norm}), i.e. assuming the worst case scenario. 

We observe that  the BNST algorithm performs significantly better as expected since the BNSL algorithm can be considered as a special case of the BNST algorithm. However, we observe that for high doppler frequencies, the BNST is not able to keep $P_{95}$ in low values, which means that the null-space changes faster than the adaptations are performed and therefore, even when SU-Tx performs a modified adaptation it causes significant interference. On the whole, we observe that for low doppler frequencies, even in this worst case scenario of independent Rayleigh Fading MIMO channels, the algorithm manages to keep the interference at low levels.

\section{Simultaneous Data Transmission And Tracking}
\label{dream}

As we have already observed, even if the SU-Tx manages to track the channel, it will be able to transmit information to SU-Rx  for a small number of time-slots and then it will need to adapt again.  Therefore, the secondary system uses a significant amount of its transmission time only to adapt to the changing environment. Motivated by this observation, the SU-Tx uses the previously presented ideas in order to satisfy the interference constraints imposed by the primary network and then we propose a transmission scheme that can be employed to transmit information superimposing the information onto the learning signals without introducing any error in the tracking.

Denote as $N = \lceil \frac{T_{FB}}{T_s} \rceil$  the number of consecutive slots that the SU-Tx transmits the same learning vector, as $\mathbf{r}_2(t)$ the information signal that SU-Tx wants to send to SU-Rx, and as $\mathbf{H}_{22}(t)$ the channel between the SU-Tx and SU-Rx.  During the $T_{FB}$ period the channels $\mathbf{H}_{12}(t)$ and $\mathbf{H}_{22}(t)$ will be approximately constant, i.e. 
\[
\mathbf{H}_{12}(t) \approx \mathbf{H}_{12}\mbox{ and }\mathbf{H}_{22}(t) \approx \mathbf{H}_{22} 
\]
\noindent because $T_{FB} << T_c$, otherwise the tracking would be impossible as we have already discussed in Section \ref{variations}.

The idea is to find a way to superimpose $\mathbf{r}_2(t) $ so that the functionality of the BNST algorithm remains unaffected and the SU-Rx receives information. We start from the signal that  the PU-Rx and the SU-Rx receive  respectively:
\begin{align}
\label{eq1}
\mathbf{y}_1(t) &= \mathbf{H}_{12} \mathbf{T} \left(\mathbf{r}_1+  \mathbf{r}_2(t)\right)+\mathbf{n}_1(t) \\
\label{eq2}
\mathbf{y}_2(t) & = \mathbf{H}_{22}\mathbf{T}(\mathbf{r}_1+\mathbf{r}_2(t))+\mathbf{n}_2(t)
\end{align}

Denote also as $\mathbf{y}_1^0(t)$ and $\mathbf{y}_2^0(t)$ the received signals at the PU-Rx and the SU-Rx respectively when no information is superimposed to the learning signal, i.e.
\begin{align}
\label{eq3}
\mathbf{y}_1^0(t) = \mathbf{H}_{12} \mathbf{T} \mathbf{r}_1+\mathbf{n}_1(t),  \mbox{  } \mathbf{y}_2^0(t) = \mathbf{H}_{22}\mathbf{T}\mathbf{r}_1+\mathbf{n}_2(t)
\end{align}
Define as 
\begin{align}
\label{eq5}
\Delta y_1 & \equiv Q(\mathbf{y}_1) -Q(\mathbf{y}_1^0)
\end{align}
the difference in the measurement at the PU-Rx as a result of the introduction of the $\mathbf{r}_2(t)$ signal, where $Q(\cdot)$ is a measurement function $Q:\mathbf{R}^N \rightarrow \mathbf{R}$ used by the PU-Rx in order to estimate the interference. The SU-Tx needs to superimpose the information over the learning signal in a way that the output of the $Q(\cdot)$ function remains unaffected. Therefore, the transmission scheme depends on the function $Q(\cdot)$. We are going to consider two different energy measurements that can be used in practice. 
\subsection{First Model of Energy Measurement}
\label{First}
\noindent ÊAssume that the PU-Rx measures the interference with the following way:
\begin{align}
\label{eq6}
Q_1(\mathbf{y}_1) = \frac{1}{N} \sum\limits_{t=1}^{N}   || \mathbf{y}_1(t)||^2
\end{align}
\noindent i.e. for each time-slot it measures the power of the interfering received signal and then it calculates the time-average over $N$ slots.

We make no assumption that the SU-Rx knows the current precoding matrix of SU-Tx nor the channel $\mathbf{H}_{22}$ between them. From equations (\ref{eq1}),(\ref{eq3}),(\ref{eq5}) it follows that
\begin{align*}
\Delta y_1 & \equiv Q_1(\mathbf{y}_1) -Q_1(\mathbf{y}_1^0) \\
 & =\frac{1}{N} \sum\limits_{t=1}^{N}  \Big ( ||  \mathbf{H_{12}} \mathbf{T} \left(\mathbf{r}_1+  \mathbf{r}_2(t)\right) +\mathbf{n}_1(t) ||^2 \\
&~~~~ - ||  \mathbf{H_{12}} \mathbf{T} \mathbf{r}_1+\mathbf{n}_1(t) ||^2  \Big )
\end{align*}
A sufficient condition for $\mathbf{r}_2(t)$ to not  affect the Jacobi method is:  $\Delta y_1 = 0$. Assume that 
\begin{align}
\mathbf{r}_2(t) & = c(t) \mathbf{r}_1
\end{align}
where $c(t) \in \mathbb{C}$ is drawn according to a known discrete probability distribution on the support $\{c_1,c_2,\cdots,c_M\}$, and $M$ is the number of different messages that can be sent. A reasonable assumption would be $Pr[c(t) = c_i] = \frac{1}{M}  ~\forall i \in \{1,2,\cdots M\}$, but we do not restrict this scheme only to this case. This assumption is the heart of the transmission scheme and it means that the SU-Tx uses the learning signal $ \mathbf{r}_1$ as the ``carrier'' signal of transmitting information. Consider the following constraints that the $\{c_i\}$ should satisfy:
\begin{itemize}
\item $\mathbb{E}\{|1 + c(t)|^2\}=1$. This is sufficient in order to ensure that the $\Delta y_1$ is close to $0$. Note that this can always be ensured if $|1+c_i|^2 = 1,\forall i \in \{1,2,\cdots M\}$, but still this is not the only way.
\item  $\mathbb{E}\{c(t)\}  \neq -1$. The necessity of this assumption will become evident later on.
\end{itemize}
It follows that
\begin{align}
\nonumber
\Delta y_1& =\frac{1}{N} \sum\limits_{t=1}^{N} \big( ||  \mathbf{H_{12}} \mathbf{T} \left[\mathbf{r}_1+  \mathbf{r}_2(t)\right]+\mathbf{n}_1(t)  ||^2 \\
\nonumber
&~~~~ -||  \mathbf{H_{12}} \mathbf{T} \mathbf{r}_1+\mathbf{n}_1(t) ||^2 \big ) \\
\label{eq34}
&\stackrel{w.p.1}{\rightarrow} \mathbb{E}\{ ||\mathbf{H}_{12} \mathbf{T}(\mathbf{r}_1 + c(t) \mathbf{r}_1)||^2\}  -  || \mathbf{H_{12}} \mathbf{T}\mathbf{r}_1||^2 \\
\nonumber
& = ||\mathbf{H}_{12} \mathbf{T}\mathbf{r}_1||^2 \mathbb{E}\{  |1 + c(t)|^2\} -  || \mathbf{H_{12}} \mathbf{T}\mathbf{r}_1||^2 \stackrel{(b)}{=}  0
\end{align}
where 
\begin{itemize}
\item  w.p.1 means with probability 1,
\item equation (\ref{eq34}) follows from the strong law of large numbers (LLN).
\item $(b)$ follows from the fact that $\mathbb{E}\{|1 + c(t)|^2\}=1$.
\end{itemize}
Thus, $\Delta y_1 \approx 0 $ for sufficiently large $N$. 

Now, we need the SU-Rx to be able to decode the received message. In the end of the $N$ slots, the SU-Rx can estimate the average of the received signal:
\begin{align}
\mathbf{\bar{y}}_2 & \equiv \frac{1}{N}\sum\limits_{t=1}^{N} \mathbf{y}_2(t)\\ 
& =  \frac{1}{N}\sum\limits_{t=1}^{N} \left ( \mathbf{H}_{22}\mathbf{T}(\mathbf{r}_1+\mathbf{r}_2(t)) +\mathbf{n}_2(t) \right ) \\
\label{eq45}
&  \stackrel{w.p.1}{\rightarrow} \mathbf{H}_{22}\mathbf{T}\mathbf{r}_1+\mathbf{H}_{22}\mathbf{T}\mathbf{r}_1(\mathbb{E}\{c(t)\}) = C \mathbf{H}_{22}\mathbf{T}\mathbf{r}_1 
\end{align}
where
\begin{itemize} 
\item equation $(\ref{eq45})$ follows from applying the LLN to both $\mathbf{n}_1(t)$ and $\mathbf{r}_2$(t) 
\item  $C=1+\mathbb{E}\{c(t)\}$ is a predefined parameter known to both the SU-Tx and SU-Rx.
\end{itemize} 

Denote as $\epsilon(N) =C \mathbf{H}_{22}\mathbf{T}\mathbf{r}_1 - \mathbf{\bar{y}}_2 $ the error introduced because $N$ is finite. Then, after $N$ slots the SU-Rx can decide which $c_i$ was sent by subtracting the value  $\frac{1}{C}\mathbf{\bar{y}}_2$ from $\mathbf{y}_2(t)$:
\begin{align}
\Delta \mathbf{y}_2 (t) & \equiv  \mathbf{y}_2 (t) -  \frac{1+c_i}{C}\mathbf{\bar{y}}_2   \\
& = (c(t)-c_i)\mathbf{H}_{22}\mathbf{T}\mathbf{r}_1 +f(\epsilon(N), \mathbf{n}_2(t))
\end{align}
\noindent where $f(\epsilon(N), \mathbf{n}_2(t)) = (1+c_i)\frac{\epsilon(N)}{C} +\mathbf{n}_2(t)$. Note that  $f(\epsilon(N), \mathbf{n}_2(t)) \rightarrow \mathbf{n}_2(t)$ for $N \rightarrow \infty$ and that $\mathbf{T}$  is not in the null space of $\mathbf{H}_{22}$  with probability 1(\cite{Noam2}).

Thus, the SU-Rx can decide amongst  which $c_i$  was sent in each time slot.  More specifically, consider the set of  hypothesis $\{H_i\}$ such that $H_i = \frac{c_i}{C} \mathbf{\bar{y}}_2$. Since $c_i$ have equal prior probability and assuming $n_2(t)$ is white Gaussian noise, the probability of error will be minimized if we follow the following decision rule:
\begin{align}
\mbox{data(t)} = {\arg\min\limits}_{1 \leq i \leq M}\{||\Delta \mathbf{y}_2(t) - H_i||\}
\end{align}

The main assumption that we have made in the above discussion is that $N$ is large enough in order the LLN to hold. Even if this is not true, we can ensure our performance requirement is met using Enumerative Coding  (Section \ref{precoding}).

\subsection{Example for a binary alphabet ($M=2$)}
\label{Example}
  
\noindent  Assume that the SU-Tx is using a $2$-constellation signal on top of $\mathbf{r}_1$
 \[
   \mathbf{r}_2(t) = \left\{ \begin{array}{rl} 
                 c_1\mathbf{r}_1, & \mbox{SU-Tx transmits 1  with prob. 0.5}  \\ c_2\mathbf{r}_1 , & \mbox{SU-Tx transmits 2 with prob. 0.5}
          \end{array} \right.
 \]
\noindent  where $c_1$ and $c_2$ are complex numbers. The constraints that the $c_1$ and $c_2$ should satisfy are:
\begin{itemize}
\item $|1 + c_1 |^2 +|1 + c_2 |^2 =2$. 
\item$c_1+ c_2\neq -2 $.
\end{itemize}
If we restrict  to the case that $c_1=e^{j\theta_0}$ and $c_2=e^{-j\theta_0}$ then both constraints are satisfied for $\theta_0 =  \frac{2 \pi}{3}$. Notice that in this specific case we always get: 
\[
||\mathbf{H_{12}} \mathbf{T}\mathbf{r}_1(1 +  e^{j\theta_0})||^2 = ||\mathbf{H_{12}} \mathbf{T}\mathbf{r}_1(1 +  e^{-j\theta_0})||^2 = ||\mathbf{H_{12}} \mathbf{T}\mathbf{r}_1||^2
\]
Then, $C = 1+\cos(\theta_0) $ and
\begin{align*}
\Delta \mathbf{y}_2 (t) & =  \mathbf{y}_2 (t)- \frac{\mathbf{\bar{y}}_2(t)}{1+\cos(\theta_0)} \\
& = \left\{ \begin{array}{rl} 
                 \frac{e^{j\theta_0}}{1+\cos(\theta_0)}\mathbf{\bar{y}}_2(t)+\mathbf{n}_2(t), & \mbox{ if 1 was transmitted}  \\\frac{e^{-j\theta_0}}{1+\cos(\theta_0)}\mathbf{\bar{y}}_2(t)+\mathbf{n}_2(t), & \mbox{if 2 was transmitted} 
          \end{array} \right.
\end{align*}
The two assumptions $H_1$ and $H_2$ of the decoder are:
\begin{align}
H_1 & = \frac{e^{j\theta_0}}{1+\cos(\theta_0)}\mathbf{\bar{y}}_2(t)  ,& H_2 & = \frac{e^{-j\theta_0}}{1+\cos(\theta_0)}\mathbf{\bar{y}}_2(t)
\end{align}
The decision rule is
\[
\mbox{data}(t) = \left\{ \begin{array}{rl} 

1, & \mbox{ if } ||\Delta \mathbf{y}_2 (t) - H_1|| <  ||\Delta \mathbf{y}_2 (t) - H_2|| \\ 2, & \mbox{other} 
          \end{array} \right.
\]

\subsection{Second Model of Energy Measurement}
\label{Second}

It is possible that the PU-Rx uses the more reliable energy measurements of the form:
\begin{align}
\label{eq9}
Q_2(\mathbf{y}_1) = ||\frac{1}{N} \sum\limits_{t=1}^{N}   \mathbf{y}_1(t)||^2
\end{align}
In this case, the proposed scheme can be modified appropriately. Again, we do not assume that the SU-Rx knows the current precoding matrix $\mathbf{T}$ nor the channel $\mathbf{H}_{22}$.  Therefore,
\begin{align}
\nonumber
\Delta y_1 & \equiv Q_2(\mathbf{y}_1) -Q_2(\mathbf{y}_1^0) \\
\nonumber
 & =|| \frac{1}{N} \sum\limits_{t=1}^{N}   (   \mathbf{H_{12}} \mathbf{T} \left(\mathbf{r}_1+  \mathbf{r}_2(t)\right) +\mathbf{n}_1(t)) ||^2 \\
\label{eq14}
&~~~~ - ||\frac{1}{N} \sum\limits_{t=1}^{N} (  \mathbf{H_{12}} \mathbf{T} \mathbf{r}_1+\mathbf{n}_1(t) )||^2 
\end{align}
Assume again as before that $\mathbf{r}_2(t) = c(t) \mathbf{r}_1$ and consider the following constraints on $c_i$:
\begin{itemize}
\item $ |1 + \mathbb{E}\{c(t)\}|^2=1$
\item Note that the constraint  $ \mathbb{E}\{c(t)\} \neq -1$ is immediately satisfied.
\end{itemize}
Starting from (\ref{eq14}) and using the LLN as before: 
\begin{align*}
\Delta y_1 & \stackrel{w.p.1}{\rightarrow} ||\mathbf{H_{12}} \mathbf{T} \mathbf{r}_1(1 +  \mathbb{E}\{c(t)\})||^2- ||\mathbf{H_{12}} \mathbf{T} \mathbf{r}_1||^2 =  0
\end{align*}
The SU-Rx in this case works exactly the same as before so we do not repeat the procedure. Note, that the only difference is that the set $\{c_i\}$ needs to satisfy different constraints.

\section{Enumerative Coding}
\label{precoding}

In the above scheme we  made the  assumption that $N$ is large enough for the LLN to hold. { To be precise, if we can ensure that
\begin{align}
\label{eq16}
\frac{1}{N}\sum\limits_{i=1}^{N}c(t) & = S
\end{align}
 for all possible transmitted sequences $c(1),c(2),\cdots, c(N)$, where $S$ is a predefined constant known to both the SU-Tx and SU-Rx, then the LLN assumption is not needed at all.} Assume the secondary system fixes $S$, then it needs a coding strategy that maps an $N$-tuple of symbols to another $N$-tuple that satisfies the constraint. 

A simple and efficient method that can ensure this constraint has been treated by Cover (\cite{Cover}). Specifically, without loss of generality, assume that the SU-Tx wants to send symbols from the alphabet $\mathcal{X}=\{1,2,\dots,M\}$  according to the uniform distribution.  This assumption models well the distribution of the actual data that are sent in a wireless network. We need some definitions from the information theoretic literature. Define the type $\mathbf{P}_{\mathbf{x}}$ of a sequence $\mathbf{x} = x_1^n$ as the relative proportion of occurrences of each symbol. Also, define as the type class of  the probability distribution $\mathbf{P}$, denoted as $T(\mathbf{P})$ the set of sequences of length $N$ that have type $\mathbf{P}_{\mathbf{x}}$, i.e. $T(P) = \{\mathbf{x} \in \mathcal{X}^N: \mathbf{P}_{\mathbf{x}} =P\}$.

For simplicity we assume that $\frac{N}{M}$ is an integer. Equation (\ref{eq16}) is satisfied if the  transmitted sequence $c(1),c(2),\dots,c(N)$ has $\mathbf{P}_c = (\frac{1}{M},\frac{1}{M},\cdots,\frac{1}{M})$, or in other words, if the secondary system uses only those sequences that belong in the type class $T(\mathbf{P}_c)$. We know that
\begin{align}
|T(\mathbf{P}_c)| = \binom{N}{\frac{N}{M},\frac{N}{M}, \dots, \frac{N}{M}}
\end{align}
Therefore, the SU-Tx can send only  $|T(\mathbf{P}_c)|$ different $N$-sequences  from the $M^N$ available sequences. For example, for $M=2$ and $N=16$ we get that $ \lfloor \log_2(|T(\mathbf{P}_c)|)\rfloor = 13 $. For M=2 and N=32,  $ \lfloor \log_2(|T(\mathbf{P}_c)|)\rfloor = 29 $ and for M=4 and N=16,  $ \lfloor \log_2(|T(\mathbf{P}_c)|)\rfloor = 25 $.

The encoding process is the following: Assume that the SU-Tx wants to transmit the sequence $x_1^n$, where $n =\lfloor \log_2(|T(\mathbf{P}_c)|)\rfloor$. Then, it treats the sequence $x_1^n$ as  the index  of an $N$-sequence that belongs to the type class $T(\mathbf{P}_c)$. Thus, it maps the  index to a sequence  and then transmits the sequence. The SU-Rx performs the reverse procedure. Both encoding and decoding procedures are more thoroughly explained in \cite{Cover} and can be easily performed with low complexity. 

\section{Simulation Results}
\label{Simulation}

\begin{figure}[h!]
\begin{center}
\includegraphics[width=0.7\columnwidth]{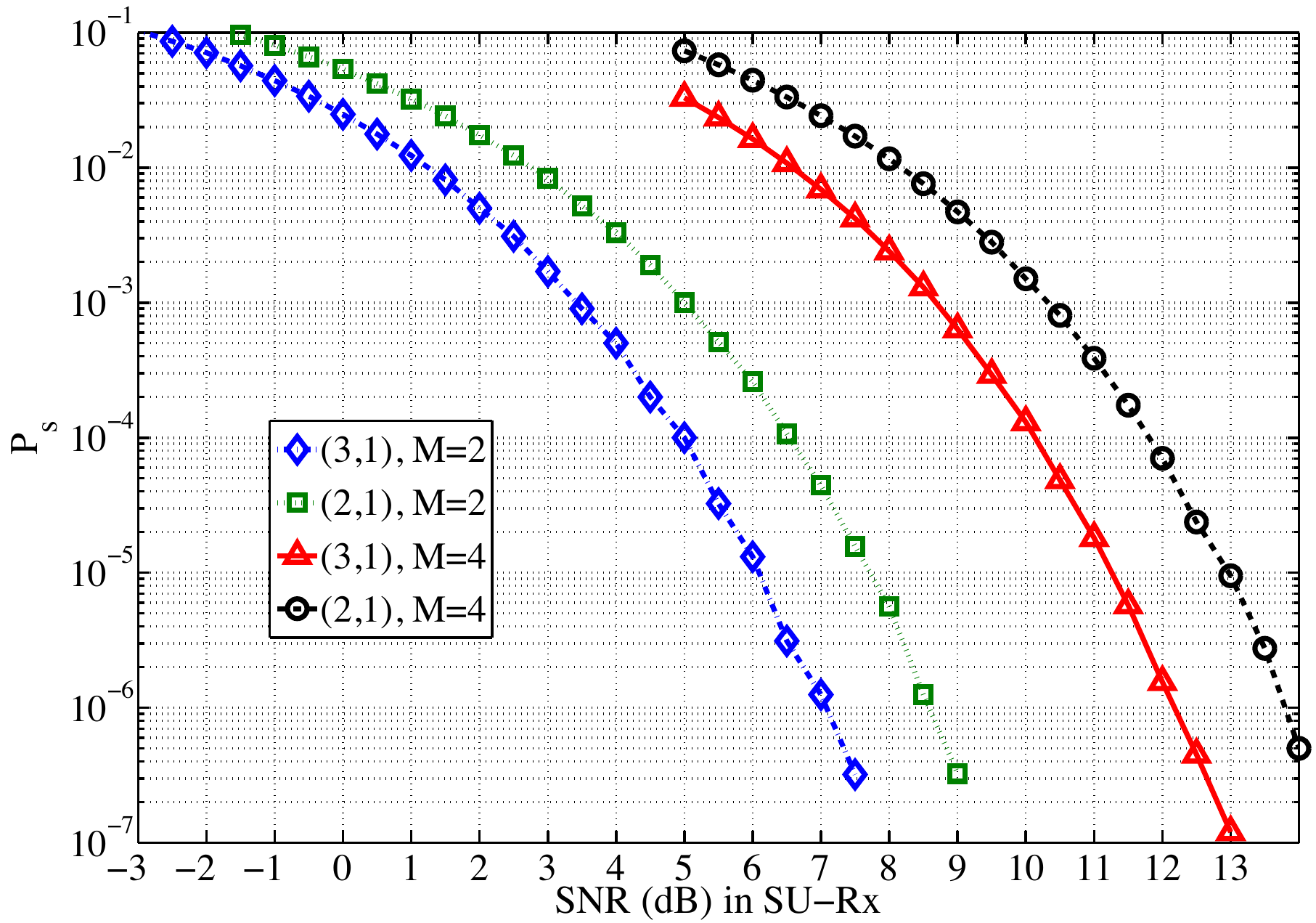}
 \caption{$P_s$ as a function of  SNR at the SU-Rx. \label{fig12}}
\end{center}
\end{figure}

In this section we test the effectiveness of the proposed scheme in its most generality, i.e. outside the BNST algorithm. Specifically, we generate $10^3$ random matrices $\mathbf{H}_{12}$, $\mathbf{H}_{22}$ and a random unitary matrix $T$. Then, for each simulation scenario we generate $10^4$ bits that are superimposed on a  vector $\mathbf{r}_1$ that was created by a random Jacobi rotation. We use $N=16$, $M=2$  (assuming $T_{FB}=1$ msec and $T_s=66.7\mu$sec) and $\theta_0=\frac{2\pi}{3}$ (See Section \ref{Example}). $P_s$ is the average probability of symbol error over all scenarios. In Figure \ref{fig12} we plot the $P_s$ as a function of the SNR at the the SU-Rx using the scheme presented in section \ref{Example}. 

The average value of $\Delta y_1$ is $0.13~dB$, which is an insignificant increase of the interference. This means that the learning procedure  remains unaffected by this transmission scheme. From Figure \ref{fig12} we observe that in the case of an $(3,1)$-system and $(2,1)$-system and $M=2$, for SNR$>3.5$ dB or SNR$>4.5$ dB respectively, the probability of error is less than $10^{-3}$. We observe that more antennas in the SU-Tx leads to a more robust system. Similar results are shown also for $M=4$.

\section{Conclusions -  Future Work}
\label{Conclusions}

The starting point of this work was the observation that the null space of a time-varying channel changes significantly faster than the coherence time of the channel. This motivated the need for a null space tracking algorithm, based on the ideas of the BNSL algorithm, which learns the null space without inflicting excessive interference  and transmits information simultaneously over the secondary system. We proposed the BNST algorithm in order to track the null space variations and showed with simulations that the proposed algorithm can actually enhance significantly the performance of the BNSL algorithm for low doppler frequencies. Lastly, we presented a transmission scheme that coexists with the learning process  by superimposing the information on the learning signal and demonstrated  its validity through simulations.

No matter which is the algorithm for null-space acquisition and/or tracking we observed through simulations that the null space under independent Rayleigh fading changes significantly faster than the coherence time. More research is needed in order to understand how fast the null space varies. Also, the Rayleigh fading MIMO channel with independent coefficients is obviously the worst case scenario since there is no LOS component. We could test the results of proposed  approach in difference channel scenarios, for example when the channel has Rician fading. 

 Last but not least, in this work we  used the vector $\mathbf{r}_1$, i.e. the current learning signal, as the carrier of the information we need to send. This gives us only one spatial degree of freedom. Is it possible to propose a similar algorithm with which we are transmitting using two or more spatial degrees of freedom and still we do not affect the learning process? These directions are going to be under investigation in our future work.

{ 

\section*{Acknowledgment}
The authors would like to thank Konstantinos Dimou for helpful discussions and suggestions. This  work was partially supported by the Center for Science of Information (CSoI), an NSF Science and Technology Center, under grant agreement CCF-0939370.

}

\end{document}